\documentclass[prd,preprint,tightenlines,floatfix,showpacs,preprintnumbers,nofootinbib,eqsecnum]{revtex4}

 \usepackage[dvips,final]{graphicx}
  \usepackage{amssymb}
   \usepackage{amsmath}
    \usepackage{amsfonts}
     \usepackage{epsfig}
      \usepackage{bm}

\usepackage[english]{babel}

\numberwithin{equation}{section}

\begin{document}

\title{Radiative decays of vector mesons\\
 in the gauge model of quark--meson interactions}

\author{V.~Beylin}
\email{vbey@rambler.ru}
\author{V.~Kuksa}%
 \email{kuksa@list.ru}
\author{G.~Vereshkov}
\email{gveresh@gmail.com} \affiliation{Research Institute of
Physics, Southern Federal University, Rostov-on-Don 344090,
Russia}
\date{\today}

\begin{abstract}

We consider meson radiative decays within the framework of
$U_0(1)\times U(1)\times SU(2)$ gauge symmetry. This approach is
based on the linear sigma-model extended by the gauge and
quark-meson interactions. Physical content and parameters of the
model are discussed. Theoretical predictions for some radiative
decays of vector mesons are in a good agreement with the experimental
data.

\end{abstract}


\pagenumbering{arabic}\setcounter{page}{1}

\maketitle
\section{Introduction}
The low energy processes with a hadron participation are permanently
in the center of theoretical and experimental activity. New
measurements and more precise experimental data force us to look for
new theoretical approaches. There are two known and widely used
methods of effective Lagrangian's approach in hadron physics: the
effective Lagrangian deriving from the QCD principles immediately
\cite{1}-\cite{11}, and the use of various dynamical symmetries to
build the basic Lagrangian in a phenomenological way
\cite{12}-\cite{18}. The linear sigma-model $(L\sigma M)$ is the
most popular and examined part of the second way. It is the main
approach to the effective analysis of nucleon-nucleon
\cite{19}-\cite{21} and quark-meson interaction \cite{12, 13, 22,
23, 24} together with the vector dominance model \cite{25}-\cite{27}.

An exact theory of the meson interactions should be nonperturbative.
That is why
 a lot of  hadron characteristics (effective couplings, formfactors etc.) are calculated, in particular,
 in the QCD Sum Rules method operating with some nonperturbative vacuum parameters.
 And that is precisely why we need in effective meson interaction theory
 giving an information on the low-energy effects and hadron structure.

An effective model of $SU_L(2)\times SU_R(2)$ chiral interactions
can be deduced from the fundamental QCD \cite{4}, using
the bosonization procedure. In such a way, the gauge structure of
meson interactions follows from the quark level QFT
\cite{4}. Thus, the initial quarks with current masses transform to
the constituent ones with effective masses $\sim 300 \,MeV$,
including some contribution of the internal gluon substructure.
Therewith, quark loops simulate an uprising of nonperturbative model
parameter. So, the proportionality of pion mass to the "quark
condensate" is restored in the framework of such effective models ~\cite{27a,27b}. Further, to incorporate
electromagnetic and strong (vectorlike) interactions into the
effective theory an appropriate vector fields are introduced as the
gauge fields. The vector meson treatment as the gauge fields realizing
some dynamical symmetry, is useful to diminish some theoretical
uncertainty of phenomenological description of the hadron interactions.
In Refs. \cite{15}-\cite{18} the gauge models were successfully used
for the consideration of some low-energy aspects of baryon-meson
interaction. It means that the fundamental quantum field principles
can be applied to the describing of interaction at the different
hierarchical levels. In the case under consideration, it provides
a transition from the quark-level sigma model ($Q \sigma M$) to
the nucleon-level sigma model ($N\sigma M$).

To construct an effective meson interaction model, we will use the
gauge scheme based on $U_0(1)\times U(1)\times SU(2)$ group. This
group is the simplest one to consider the light unflavored vector mesons, $\rho$ and $\omega$,
together with the photon, as the gauge fields.
Moreover, it has the necessary symmetry to describe the
electromagnetic and strong meson-meson and quark-meson interactions,
which are insensitive to the chiral structure. So, to analyze the
electromagnetic and strong effects only, it is sufficient to
localize the diagonal sum of the global chiral $SU_L(2)\times
SU_R(2)$. This sum corresponds to $SU(2)$ subgroup of the total gauge
group $U_0(1)\times U(1)\times SU(2)$. The extra $U(1)$ groups are
introduced for the gauge realization of the vector meson dominance
(see the next section). Note that in the model we deal with the
constituent quarks not with the fundamental current ones.
Obviously, the electromagnetic interaction violates the isospin
symmetry.

After the spontaneous breaking of
the symmetry, Higgs fields can be associated with the
scalar mesons $a_0(980)$ and $f_0(980)$ (see the next section).

As it was noted above, this model reproduces the relation
$m^2_{\pi}\sim m_q <\bar q q>$, if the quark condensate is taken into
account in the equation for the vacuum shift. Further, quarks arise in
the model as the gauge group representations, and these degrees of
freedom account for the internal meson structure. In fact,
quark-meson models (see also \cite{22,28}) are necessary hybrid
approaches to reproduce both nonperturbative and structure effects
of the low-energy meson interactions.

In this paper, the gauge scheme interactions of $\omega$, $\rho$ and
$\gamma$ with quarks and mesons are analyzed. In particular, this
scheme describes some important features of radiative decays of
vector mesons. Namely, the electromagnetic quanta mixing with
$\omega$- and $\rho$- fields leads to the gauge variant of vector
dominance, i.e. all these interactions are considered in the same
way. So, we deal with the $Q\sigma M$ which incorporates gauge
interactions of vector mesons with the constituent quarks. Then, the model accounts
for hadronic internal degrees of freedom, which are described in the
quantum field approach. Really, two-level (quark-meson)
structure of the model realizes some analog of the "bootstrap" idea.

The model contains a number of free parameters in the gauge sector,
which can be fixed from the well measured two-particle decays of vector mesons.
As it will be shown later, the gauge model
predictions for the radiative vector meson decays are strict (without any free parameters).

As a rule, the radiative decays $\rho^0 \to \pi^+ \pi^-\gamma$ and
$\omega \to \pi^+ \pi^-\gamma$ were described in a phenomenological
way. Within the framework of the gauge model, these processes arise at
the tree level. There is a good agreement between the theory and the
experimental data on the vector meson radiative decays.
So, these
decays can be successfully studied in the gauge field approach.

Internal meson structure is described by the quark-meson sector of
the model. These aspects of the model have been tested in $\rho \to
\pi^0 \gamma$, $\omega \to \pi^0 \gamma$ and $\rho^0\to e^+e^-$
decays and results of calculations are in a good agreement with the
experimental data. Analysis of the decay $\omega \to \pi^0 \pi^+
\pi^-$ \cite{28,29a,29b,29c}, which arises at the loop level due to the
quark-meson interaction only, is more cumbersome. Calculations done show that the three-pion decays of $\omega-$ and $\rho-$ mesons
are well descripted in this model. The obtained results are in preparation now.

The structure of the paper is as follows. In Section 2 the gauge
field model is described. Numerical results for tree decays
$\rho^0\rightarrow\pi^+\pi^-\gamma$ and
  $\omega\rightarrow\pi^+\pi^-\gamma$ are presented in Section 3.
We consider loop decays $\rho^0\rightarrow\pi^0 \gamma$, $\omega\rightarrow\pi^0 \gamma$ and $\rho^0\to e^+e^-$ in Section 4.

\section{The gauge field model of meson-meson and quark-meson interactions}

The gauge model of the low-energy meson interactions is based on the
conceptual assumption concerning the transfer of gauge principles
from the fundamental quark level to the effective hadron one. Just
the same transfer was deduced in Ref. \cite{4} for the case of
quark-meson hierarchy levels. The applicability of the gauge
approach to the baryon-meson interaction was demonstrated in Refs.
\cite{15}-\cite{18}. In this work, we show the validity of the gauge
scheme for the case of meson-meson and quark-meson effective
interactions.

We describe $\omega$- and $\rho$- mesons together with $\gamma$-
field within the framework of the gauge approach. Thus, the strong and
electromagnetic interactions are united in the gauge scheme. The
simplest variant of corresponding dynamical symmetry is based on
$U_{0}(1)\times U(1)\times SU(2)$ group. As it must be, the global
symmetry of the chiral model takes place here, and, in a general case,
the total chiral group of symmetry should be localized.
However, for the application of the gauge approach to the $\rho-$ and $\omega-$ strong and electromagnetic decays,
it is reasonably to localize the diagonal sum of $SU_{L,R}(2)$ subgroups together with the additional $U(1)$ groups.

The traditional $\sigma$- model is a part of the model involving
nonlinear terms of self-action. The $\sigma$-model symmetry can be
realized in various representations of the global groups: $O(4),
\,\,SU_L(2)\times SU_R(2), \,\,SU(2)\times G(3)$, where $SU(2)$
means the diagonal sum of $SU_{L,R}(2)$ subgroups.
 For all these cases we have six independent parameters of transformations.
Transformation of the quark doublet is:
\begin{align}\label{2.0}
q^{'}&=q+\frac{i}{2}(\alpha_a\tau^a+\beta_a\tau^a\gamma_5)q\notag\\
   &=q+\frac{i}{2}(\alpha+\beta)\tau^a q_R+\frac{i}{2}(\alpha-\beta)_a\tau^a q_L,
\end{align}
where $\alpha_a$ are the parameters of the $SU_{L+R}(2)$ group and $\beta_a$ are the parameters of
the  $SU_{L-R}(2)$ group.
Thus, here we use $SU(2)\times G(3)=SU_{L+R}(2)\times SU_{L-R}(2)$ representation for $\pi$ and $\sigma$ fields.
Then, the corresponding transformation properties fo $(\sigma,\pi)$ are:
\begin{align}
\pi^{'}_a=&\pi_a+\epsilon_{abc}\alpha_b \pi_c+\beta_a\sigma,\notag\\
\sigma^{'}=&\sigma+\beta_a\pi_a\notag.
\end{align}

Having two sets of independent parameters $\alpha_b$ and
$\beta_a$, to describe strong and electromagnetic interactions it
is sufficient to localize the $\alpha_b$ set together with the
parameters of $U(1)$ groups. (To consider weak interactions the
localization of the $\beta_a$ parameters should be added.)
As it is seen,
 the triplet of pions is the adjoined representation of the
gauge group. At the same time, the quark doublet $u, \,d$ is the
fundamental representation of this group. Thus, we consider the
triplet $\pi_a$, the singlet $\sigma$-meson and $u, d$-quarks as
fields of matter.

The physical scalar fields, emerging from the initial Higgs
multiplets, can be associated with the scalar mesons $a_0(980)$
and $f_0(980)$.

From these considerations the model Lagrangian is:
\begin{align}\label{E:8}
L &=i \bar{q} \hat{D} q - \varkappa \bar{q} (\sigma +i\pi^a\tau_a \gamma_5)q + \frac{1}{2}(D_{\mu}\pi^a)^+(D_{\mu}\pi^a)+\frac{1}{2}\partial_{\mu}\sigma
\partial^{\mu}\sigma + \frac{1}{2}\mu^2(\sigma^2+\pi^a\pi^a)\notag \\
&-\frac{1}{4}\lambda(\sigma^2+\pi^a\pi^a)^2+(D_{\mu}H_A)^+(D_{\mu}H_A)+
\mu_A^2(H_A^+H_A)-\lambda_1(H_A^+H_A)^2-\lambda_2(H_A^+H_B)(H_B^+H_A)\notag\\
& -h(H_A^+H_A)(\sigma^2+\pi^a\pi^a)-\frac{1}{4}B_{\mu
\nu}B^{\mu \nu}-\frac{1}{4}V_{\mu \nu}V^{\mu
\nu}-\frac{1}{4}V^a_{\mu \nu}V_a^{\mu \nu}.
\end{align}
Here $q = (u,d)$ - is the first generation quark doublet;
$H_{1,2}$ - two scalar fields doublets with hypercharges
$Y_{1,2}=\pm 1/2$, $a=1,2,3$ and $A=1,2$. The gauge derivatives and field strengthes are:
\begin{align}
&\hat{D}q = \gamma^{\mu}(\partial_{\mu}-
\frac{i}{6}g_{0}B_{\mu}-\frac{i}{2}g_1 V_{\mu}-\frac{i}{2}g_2 V^a_{\mu}\tau_a)q;\notag\\ &D_{\mu}\pi_a=\partial_{\mu}\pi_a-ig_2 V^b_{\mu}\epsilon_{bac}\pi_c;\notag\\
&D_{\mu}H_{1,2} = (\partial_{\mu}\pm
\frac{i}{2}g_{0}B_{\mu}-\frac{i}{2}g_1 V_{\mu}-\frac{i}{2}g_2\tau_a
V^a_{\mu})H_{1,2};\notag\\
&B_{\mu\nu}=\partial_{\mu}B_{\nu}-\partial_{\nu}B_{\mu},\,\,\,V_{\mu\nu}=
\partial_{\mu}V_{\nu}-\partial_{\nu}V_{\mu};\notag\\
&V^a_{\mu\nu}=\partial_{\mu}V^a_{\nu}-\partial_{\nu}V^a_{\mu}+g_2\epsilon ^{abc}V^b_{\mu}V^c_{\nu}.
\label{E:9}
\end{align}
In principle, the gauge invariant term $B_{\mu \nu}V^{\mu \nu}$ can be added to the Lagrangian.
However, this term can be diagonalized from the very beginning by the orthogonal
transformation of the initial $B_{\mu}$ and $V_{\mu}$ fields. So, we get the same Lagrangian with the redefined $B_{\mu}$ and $V_{\mu}$ fields.

To describe the realistic processes of decays and scattering we
introduce also the interactions of leptons with $U_{0}(1)$ and
$U(1)$ fields in the gauge  form:
\begin{equation}
L_l=i\bar l \hat D l=i \bar l \gamma^{\mu}(\partial _{\mu} -
ig_{0}B_{\mu} -i\varepsilon g_1V_{\mu})l,
\label{E:10}
\end{equation}
where interaction of $V_{\mu}$ field with leptons is driven by the
phenomenological coefficient $\varepsilon$. Note that the origin of
the last term in the above
 formula is specific. Namely, this term is caused by the nondiagonal $B_{\mu \nu}V^{\mu \nu}$
 part of the initial Lagrangian. Thus, the interaction of leptons with the $V_{\mu}$ vector field
 occurs allowing to describe the decays $\rho^0\to
e^{+}e^{-},\,\,\,\omega\to e^{+}e^{-}$  (see the Section 4).

The introducing of the vector fields to the theory in a gauge way
provides universality of coupling with vector fields, i.e. it strictly
bounds the number of free parameters. So, this approach raises the
predictability of the model. The analogous universality of vector
and pseudoscalar meson interactions was analyzed in \cite{24} in a
phenomenological way.

Physical states are formed by the primary fields mixing when mass quadratic
forms of scalar and vector fields are diagonalized. At the tree level the mass forms arise
as a result of vacuum shifts:
$$<\sigma>=v,\,\,\, <H_1>=\frac{1}{\sqrt{2}}(v_1,0),\,\,\,
<H_2>=\frac{1}{\sqrt{2}}(0,v_2).$$

Note, the loop contributions
to the mass matrix can play an essential role in the mixing of mesons.
So, in a general case this matrix has the form:
$$M^2(s)=M^2_0+\Pi(s),$$
where $M^2_0$ is formed by the shifts and $\Pi(s)$ by the self-energy insertions.
In our case these insertions are significant for the description of the $\omega-\rho^0$ mixing.

As the result of the diagonalization of vector and scalar mass
forms, we get the model spectrum of the vector and scalar multiplets.
The gauge status of vector mesons $\rho(770)$ and $\omega(782)$ is
confirmed by the calculation of their model mass spectrum and decay
properties. These model parameters are in agreement with the
experimental data for dominant decay channels of vector mesons ~\cite{29d}.
 Moreover, due to the presence of free parameters
in the scalar sector, it is possible to describe the mass spectrum and decay channels of
the scalar mesons $\sigma$ (or $f_0(600)$), $a_0(980)$ and $f_0(980)$,
which have the status of Higgs scalars in the model (see also
Eqs.(\ref{E:14}) and comment).

An emergence of primary hadron component in the photon is an
important consequence of the mixing in the gauge sector. The structure
of the vector boson physical states can be illustrated in a tree approximation (when $M^2(s)=M^2_0$)
by the following expressions:
\begin{align}\label{E:11}
&A_{\mu}=\cos\theta \cdot B_{\mu}+\sin\theta \cdot V^3_{\mu}, \notag\\
&\omega_{\mu} = \cos\phi \cdot V_{\mu}+\sin\phi \cdot
(\sin\theta\cdot B_{\mu}-\cos\theta \cdot V^3_{\mu}),\notag\\
&\rho^0_{\mu} = \sin\phi \cdot V_{\mu}+\cos\phi \cdot
(-\sin\theta\cdot B_{\mu}+\cos\theta \cdot V^3_{\mu}),
\end{align}
where the mixing angle $\theta$ is determined under the
diagonalization of the vector fields quadratic form. Due to the mixing
(\ref{E:11}) the processes with initial "unphysical" photon,
$e^+e^-\to \gamma '\to X$, give contributions to the processes with  the intermediate hadron states:
\begin{equation}\label{E:12}
e^+e^-\to \gamma,\omega, \rho^0\to X.
\end{equation}

 Some parameters of the mixing can be fixed from the experimental data on the vector meson decays.
 It will allows to describe some decays properties of vector mesons
considered in this paper.
The tree form of the mixing (\ref{E:11}) is caused by the diagonalization of the real mass matrix
(this matrix is generated by the vacuum shifts only, without an account of any self-energy insertions)
with the help of the real orthogonal matrix. Together with the tree real gauge couplings,
it leads to the absence of the relative phase in the $\rho \pi \pi$ and $\omega \pi \pi$ channels at the tree level (see (\ref{E:12a})).
Experiments, however, indicate that there is a nonzero relative phase shift between amplitudes
of $e^+e^- \to \rho^0 \to \pi^+ \pi^-$ and $e^+e^- \to \omega \to \pi^+ \pi^-$ processes
(it is interpreted as $\rho -\omega$ mixing effect in the pion formfactor).

In a phenomenological approach, the complexity is introduced into
the superposition of pure isospin $|\rho_I>$ and $|\omega_I>$
states. So, the needed relative phase is a free parameter which is
determined from the fit to the experiment
~\cite{29e,29f,29g,29h,29j,29k}.

 Note, such
 important feature of vector mesons physics
as the $\rho^0 - \omega$ mixing, can be reproduced in the model at the loop level.
 As a rule, physical vector states of $\rho-$ and $\omega-$ mesons are described as linear combinations of pure isospin states with complex coefficients. It results to the relative $\rho- \omega$ mixing phase.
 In our model the complexity cannot be directly introduced into the superposition of operator fields (\ref{E:11}), because it describes the physical neutral fields. To reproduce the relative phase, we would treat the complexity in a following way. Namely, instead of tree $\sin \phi$ and $\cos \phi$ we should have absolute values of the renormalized parameters of mixing in the superposition (\ref{E:11}). Then, the corresponding phase should be introduced into the renormalized couplings as the vertex factors (see (\ref{E:12a}) and (\ref{E:12b})). This treatment is equivalent to the known approach when physical meson states are the complex superpositions of pure isospin vector states ~\cite{29e,29f,29g}.

The necessary phase value can be provided in the framework of this model due to presence of free parameters,
in particular, in the scalar sector ($\sigma,\, f_0,\, a_0,\,\pi-$ couplings). Certainly,
these free parameters should be fixed simultaneously from the analysis of the $\sigma, \, f_0, \, a_0$- decays.

For the processes, which are under consideration in this paper, the relative $\rho-\omega$ phase is inessential, so we do not use it here.
An analysis of the vector mesons mass matrix with an account of loop contributions and the study
of the $\rho - \omega$ mixing in the model will be the subject of a forthcoming paper.

Now, we give the main part of the physical Lagrangian which will be used for calculations:
\begin{align}\label{E:12a}
 L_{Phys}&=\bar{u}\gamma^{\mu}u(\frac{2}{3}eA_{\mu}+g_{u\omega}\omega_{\mu}+g_{u\rho}\rho^0_{\mu})+
          \bar{d}\gamma^{\mu}d(-\frac{1}{3}eA_{\mu}+g_{d\omega}\omega_{\mu}+g_{d\rho}\rho^0_{\mu})\notag\\
         &+\frac{1}{\sqrt{2}}g_2 \rho^{+}_{\mu}\bar{u}\gamma^{\mu}d+\frac{1}{\sqrt{2}}g_2 \rho^{-}_{\mu}\bar{d}\gamma^{\mu}u
          +ig_2\rho^{+\mu}(\pi^0\pi^{-}_{,\mu}-\pi^{-}\pi^0_{,\mu})+
          ig_2\rho^{-\mu}(\pi^{+}\pi^0_{,\mu}-\pi^0\pi^{+}_{,\mu})\notag\\
         &+ig_2(\pi^{-}\pi^{+}_{,\mu}-\pi^{+}\pi^{-}_{,\mu})(\sin \theta \,\, A^{\mu}-\cos \theta s_{\phi} \,\,\omega^{\mu}+\cos \theta c_{\phi} \,\,\rho^{0\mu})\notag\\
         &-\sqrt{2}i\varkappa\pi^{+}\bar{u}\gamma_5 d-\sqrt{2}i\varkappa\pi^{-}\bar{d}\gamma_5 u-i\varkappa\pi^0(\bar{u}\gamma_5 u-\bar{d}\gamma_5 d)\notag\\
         &+2g_2e\cos \theta c_{\phi} \,\,\rho^0_{\mu}A^{\mu}\pi^{+}\pi^{-}-
          2g_2e\cos \theta s_{\phi} \,\,\omega_{\mu}A^{\mu}\pi^{+}\pi^{-}.
\end{align}
In Eqs.(\ref{E:12a}):
\begin{align}\label{E:12b}
 &g_{u\omega}=\frac{1}{2}g_1 c_{\phi} +\frac{1}{2}s_{\phi} \,(\frac{1}{3}g_0\sin \theta-g_2\cos \theta)  \,,\notag\\ &g_{u\rho}=\frac{1}{2}g_1 s_{\phi} -\frac{1}{2}c_{\phi} \,(\frac{1}{3}g_0\sin \theta-g_2\cos \theta)\,, \notag\\
 &g_{d\omega}=\frac{1}{2}g_1 c_{\phi} +\frac{1}{2} s_{\phi} \,(\frac{1}{3}g_0\sin \theta+g_2\cos \theta) \,,\notag\\ &g_{d\rho}=\frac{1}{2}g_1 s_{\phi} -\frac{1}{2} c_{\phi} \,(\frac{1}{3}g_0\sin \theta+g_2\cos \theta)\,,
\end{align}
where $s_{\phi}$ and $c_{\psi}$ are the complex (renormalized) parameters of the $V-B$ mixing with account of the self-energy insertion to the mass matrix.

There are also another parts of the interaction Lagrangian,
which have the structures $HVV$ and $HHV$. Here $H$ and $V$ are the sets of scalar and vector physical fields.
The used identification of the gauge and Higgs fields components provides the invariance of the physical Lagrangian under the $CP$ transformations.

Some relations arise as a direct consequences of the model:
 \begin{equation}\label{E:12c}
  \sin \theta=\frac{g_{0}}{\sqrt{g_0^2+g_2^2}},\,\,\,e=g_0\cos \theta,\,\,\, v^2_1+v^2_2=4 \frac{m^2_{\rho^{\pm}}}{g^2_2},\,\,\,
  |s_{\phi}|=\frac{g_1}{g_2}\sqrt{\frac{m^2_{\rho^{\pm}}-m^2_{\omega}(g^2_2/g^2_1)}
  {m^2_{\omega}-m^2_{\rho^0}}}.
 \end{equation}

  The value of $g_{0}$ can be found from the second relation $e=g_{0}\cdot
 g_2/(g_{0}^2+g_2^2)^{1/2}$. Values of $g_2$ and $|s_{\phi}|$ (or $g_1$) are
fixed from the experimental values of $\Gamma
 (\rho^+ \to \pi^+ \pi^0)$ and  $\Gamma
 (\omega \to \pi^+ \pi^-)$. It should be noted, the last decay
 takes place due to the mixing (\ref{E:11}) only. Thus, we extract the values of main model parameters
 which we use to describe another decay properties of mesons:
 \begin{align}\label{E:13}
 g_{0}^2/4\pi = 7.32\cdot 10^{-3}, \,\,\,
 g_1^2/4\pi= 2.86, \,\,\, g_2^2/4\pi = 2.81,\notag\\
 |s_{\phi}| = 0.031, \, \sin \theta = 0.051,\, v_1^2 + v_2^2\approx
 (250.7 \,\mbox{MeV})^2.
 \end{align}
 These values were applied for calculation of the vector meson radiative decay widths to verify the gauge vector dominance approach.
 In our strategy of calculations the strong couplings are extracted from the above mentioned processes as the effective final values.
So, we do not need in loop corrections to these couplings. At the same time, electromagnetic vertices should be renormalized
 by the strong interactions (for details, see the decay $\rho^0\to e^+e^-$ which is considered in the fourth section).

 In addition to (\ref{E:12a}) we give also a part of the Lagrangian, describing an interaction of the scalar mesons with $\pi$-mesons:
 \begin{equation}\label{E:14}
  L_{\pi h}=(\pi^0\pi^0+2\pi^{+}\pi^{-})(g_{\sigma\pi}\sigma_0+g_{f\pi}f_0+g_{a\pi}a_0).
 \end{equation}
 This part of the Lagrangian contains free coupling constants and makes it possible to describe the dominant decay channels of scalar mesons -
  $f_0(980)\to\pi\pi$ and $\sigma_0\equiv f_0(600)\to \pi\pi$ ~\cite{29d}.
The decay channel $a_0\to\pi\pi$ is not observed and from the model coupling
\[g_{a\pi}=h(v_2-v_1)/\sqrt{2}\] it follows that $v_2\approxeq v_1$.
So, in the scalar sector the residual global $SU(2)$- symmetry approximately takes place after the shift.
Exact symmetry, $v_1=v_2$, is not possible because it forbids the decay channel $\omega\to\pi^{+}\pi^{-}$, which is observable experimentally.
  Evidently, the inequality $v_1\ne v_2$, violating the global isotopic $SU(2)$ symmetry,
is related with an account of the electromagnetic interaction in the model.

 There are no two-photon decays of scalar mesons at the tree level, i.e. these channels are suppressed in the frame of the model.
Thus, the status of scalars as the Higgs mesons does not contradict to their observed mass spectra and decay properties.

 Because of the model two-level structure corresponding to the meson-meson and quark-meson interactions,
 both of them are tested independently in two-particle and three-particle channels. It is seen from (\ref{E:12a})
 that the meson-meson Lagrangian describes both tree and loop level processes,
while the quark-meson interactions occur in the model at the loop level only.

  \section{Radiative decays $\rho^0\rightarrow\pi^+\pi^-\gamma$ and
  $\omega\rightarrow\pi^+\pi^-\gamma$}

 Radiative decays of neutral vector mesons of the type $V\to \pi \pi \gamma$ are an object of steady attention during some
 decades (see \cite{29a,29b} and also \cite{30}-\cite{34}). Experimental investigation of these reactions and accompanied
 theoretical speculations
 contribute to the understanding of hadron
intermediate states and low
 energy dynamic of meson interactions.

In the channels where the
 charged pair $\pi^+ \pi^-$ is formed, dominant contribution comes from tree diagrams
 corresponding to the vector dominance approximation \cite{31,34}.
Radiative decays $\rho^0 \to \pi^+\pi^-\gamma$ and $\omega \to
 \pi^+\pi^-\gamma$ are described by meson-meson sector of the gauge model.

 At the tree level, the former decay is represented by the diagrams in
 Fig.1.
\begin{figure}[h!]
\centerline{\epsfig{file=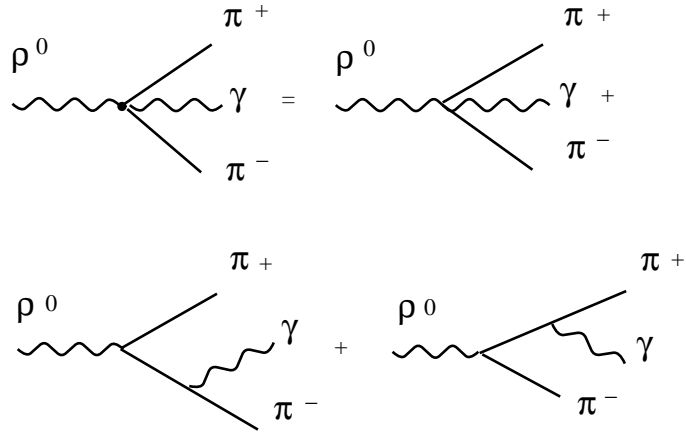,width=9cm}}
\caption{Feynman diagrams for radiative decay $\rho^0\to \pi^+
\pi^- \gamma$.} \label{fig:Feynm1}
\end{figure}
 The total amplitude for the process is:
 \begin{equation}\label{E:1}
  M^{tot} =\frac{i g e^{\mu}_{\rho}e^{\nu}_{\gamma}}{8
 \pi^2(k^0_{\rho}k^0_{\gamma}k^0_+k^0_-)^{1/2}}\left[g_{\mu \nu} +
 \frac{2k^-_{\mu}k^+_{\nu}}{(k_{\gamma}+k_+)^2-m^2_{\pi}+}+
 \frac{2k^-_{\nu}k^+_{\mu}}{(k_{\gamma}+k_-)^2-m^2_{\pi}}\right].
 \end{equation}
Here $g=e g_2 \cos\theta |c_{\phi}|$ and $e^{\mu}_{\rho}, \,
e^{\nu}_{\gamma}$ are polarization vectors for $\rho^0$ - meson and
photon, $k_{\rho}, \, k_{\gamma}, \, k_+, \, k_-$ are 4-momenta for
all particles in the process. In (\ref{E:1}) we omit all terms which
are equal to zero on the mass shell in the transversal gauge. For
comparison with the experimental spectrum of photons (see the work
of Dolinsky, \cite{33}) the differential width is presented in the form:
\begin{equation}\label{E:2}
d\Gamma(E_{\gamma})/dE_{\gamma} =
\frac{G}{\kappa}\left(F_1(\kappa)+F_2(\kappa) \ln F_3(\kappa)\right),
\end{equation}
where:
\begin{align}\label{E:3}
\kappa &= E_{\gamma}/m_{\rho}, \,\,\, G = \alpha_{em}\cdot g_2^2
\cos^2\theta \cdot |c_{\phi}|^2/24 \pi^2,\,\,\, \mu
=m_{\pi}^2/m^2_{\rho}\,, \notag\\
&F_1(\kappa)=\left(\frac{1-2\kappa-4\mu}{1-2\kappa}\right)^{1/2}
\left(-1+2\kappa+4\kappa^2+4\mu(1-2\kappa)\right);\notag\\
&F_2(\kappa)=1 -2\kappa-2\mu(3-4\kappa-4\mu);\notag\\ &F_3(\kappa)
=\frac{1}{2\mu}\left[1-2\kappa-2\mu+\left((1-2\kappa)\cdot (1-2\kappa -
4\mu)\right)^{1/2}\right].
\end{align}

\begin{figure}[h!]
\centerline{\epsfig{file=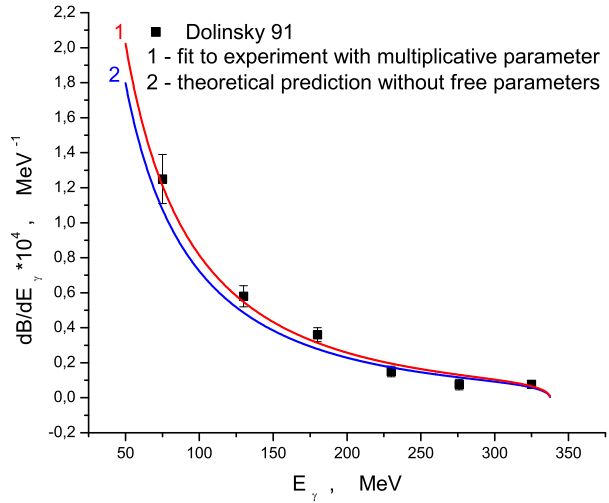,width=9cm}}
\caption{Photons spectrum in $\rho\to 2\pi\gamma$ decay .}
\label{fig:Curve}
\end{figure}

 Our numerical results, which follow from (\ref{E:2}) and (\ref{E:3}), agree with the results of
~\cite{30,31,32} given by the vector dominance approach. In Fig.2
the theoretical spectrum of photons in comparison with the
experimental data (from ~\cite{33}) are represented. The curve (2)
in Fig.2 describes the spectrum $dB(E_{\gamma})/dE_{\gamma}$,
normalized by the total width. Here
$dB(E_{\gamma})=d\Gamma(E_{\gamma})/\Gamma_{tot}^{\rho}$, and it
does not depend on the model coupling constants. The curve (1)
represents the model fit of the experimental data with the help of a
single free multiplicative parameter. As it is seen, this fitted
curve (1) improves the prediction at the low energy range. This
effect can be caused by the renormalization of $g_{\rho \pi \pi}$
coupling.

To
improve the theoretical spectrum near $E_{max}$ it was suggested to consider
the loop corrections (see \cite{30} and \cite{31}).
However, this consideration is reasonable only if the more detailed and
reliable experimental data on the photon spectrum are available.

Integration of (\ref{E:2}) from $E_{\gamma}^{min} = 50 \,\mbox{MeV}
$ up to $E_{\gamma}^{max} = m_{\rho}(1-4\mu)/2$ gives the value of
partial $\rho$- meson branching $B(\rho^0\to \pi^+ \pi^- \gamma) =
1.17 \cdot 10^{-2}$ which slightly exceeds the experimental value
$B^{exp}(\rho^0\to \pi^+ \pi^- \gamma) = (0.99 \pm 0.16)\cdot
10^{-2}$ from ~\cite{29d}. An account of
the loop contributions (with the phenomenological couplings) leads to the result:
$B^{phen}(\rho^0\to \pi^+ \pi^- \gamma) = (1.22 \pm 0.02)\cdot
10^{-2}$ ~\cite{31}. It should be noted that there is some
discrepancy between the experimental branching and the one following
from the integration of spectrum (see Fig.2). Namely, an excess of
$B^{theor}$ over $B^{exp}$ can be caused by the deviation of the
theoretical spectrum from the experimental one at the energy range
$E_{\gamma}<75 \mbox{MeV}$ (to the left of the first point of the
experimental data in Fig.2).

Decay characteristics of the process $\omega\to \pi^+ \pi^-\gamma$ at the tree level
are computed analogously with the following replacement in
(\ref{E:1}) - (\ref{E:3}): $ c_{\phi} \to s_{\phi}$ in $G$, and
$m_{\rho} \to m_{\omega}$ in $\kappa$ and $\mu$. The partial width
for the decay is damped by the small mixing parameter, $|s_{\phi}|
\approx 0.034$, so we have $B(\omega\to \pi^+ \pi^- \gamma) = 4.0
\cdot 10^{-4}$ and $B(\omega\to \pi^+ \pi^- \gamma) = 2.6 \cdot
10^{-4}$ for $E^{min}_{\gamma} = 30 \, \mbox{MeV}$ and $50\,
\mbox{MeV}$, respectively. These estimations do not contradict to
the experimental restriction $B^{exp}(\omega\to \pi^+ \pi^- \gamma) \le 3.6
\cdot 10^{-3}$ ~\cite{29d} and agree with the theoretical
results of ~\cite{34}. Thus, the same set of the fixed parameters (the gauge constant and two mixing tree angles)
describes two different decays at the tree level.
Certainly, loop corrections can be important for the case due to the smallness of the
tree contribution, see, for example ~\cite{34,34a}. In analogy with the results of these papers,
loop corrections can increase $B(\omega\to \pi^+ \pi^- \gamma)$ up to $(2-3) \cdot 10^{-3}$,
which does not contradict again to the upper limit of the experiment.

As it was noted above, consideration of the loop corrections to the decay $\rho^0\to \pi^+ \pi^-
\gamma$ in \cite{31}-\cite{34} was intended to describe the photon
spectrum fine structure. However, it increases the discrepancy
between the model and experimental values of the total width.
Quantitatively, this effect depends on the underlying model. Moreover,
in effective theories an account of loop diagrams (for some process
arising at the tree level) has some subtleties connected with
the compensation of divergencies and renormalizability. For the
processes which occur at the loop level only, these problems are
absent --- all divergencies are summed to zero when all external
lines are on the mass shell ~\cite{36}. An examples of such loop
processes will be given in the next section.

The usage of parameters from (\ref{E:13})
together with the constants $g_{\rho ee}$ and $g_{\omega ee}$
extracted from the widths of $\rho^0 \to ee$ and  $\omega \to ee$
decays, leads to the correct estimation of the cross section in the peaks vicinity,
$\sigma^{theor} \approx \sigma^{exp}\approx 1.3 \,\mbox{mkb}$.

As it was noted above, the tree approximation (without an account of the $\rho -\omega$ mixing at the loop level) does
not allow us to describe the resonance curve in this region in details.
A relative $\rho-\omega$ phase consideration is a very essential for the resonance curve description.

\section{Radiative decays $\rho^0\rightarrow\pi^0\gamma$, $\omega\rightarrow\pi^0\gamma$ and leptonic decay $\rho^0\rightarrow e^+e^-$}

In the most papers these decays are defined by the phenomenological
vertices which are introduced at the tree level \cite{1}. In the
model considered, radiative decays $\rho^0,\, \omega \to
\pi^0\gamma$ and three-particle decays $\omega, \rho \to 3 \pi$ occur via
quark loops with the gauge vertices. One-loop diagrams for the
decays $\omega,\, \rho^0 \to \pi^0\gamma$ are shown in Fig.3, where
$q = u,d$ denote quark fields.
\begin{figure}[h!]
\centerline{\epsfig{file=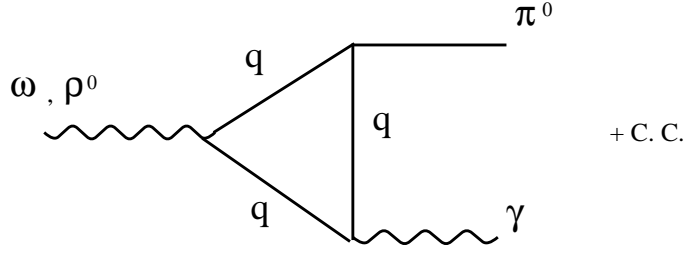,width=9cm}}
\caption{Feynman diagrams for the radiative decay $\rho \to\pi^0
\gamma$.} \label{fig:Feynm2}
\end{figure}

Total amplitude for  the
process $\omega \to \pi^0\gamma$ has the following form
\begin{equation}\label{E:4}
M_{\omega}=\frac{-2i\pi^2 N_cgm_q}{(2\pi)^{9/2}(2p^0k^0_{\gamma}k^0_{\pi})^{1/2}}
e^{\mu}_{\omega}e^{\nu}_{\gamma}k^{\alpha}_{\gamma}p^{\beta}\epsilon_{\mu
\nu \alpha \beta} \cdot C_0(0,m_{\omega}^2,m_{\pi}^2;m_q,m_q,m_q).
\end{equation}
Here $N_c = 3$ is the color factor, vertex constant $g = g_1e
\varkappa |c_{\phi}|$ ($\varkappa$ is the constant of
$qq\sigma$-interaction, see (\ref{E:12a})) and $C_0(0,
m_{\pi}^2,m_{\omega}^2;m_q,m_q,m_q)$ is the three-point Passarino -
Veltman function \cite{37}. For the constituent quark mass we
suppose $m_u \approx m_d = m_q$. Substituting the function $C_0$
into (\ref{E:4}), we have:
\begin{equation}\label{E:5}
\Gamma(\omega \to \pi^0 \gamma) =\frac{3\alpha
g_1^2}{2^7\pi^4}|c_{\phi}|^2\,\,
m_q\frac{m_q^3}{m_{\omega}f_{\pi}^2}\left(1-\frac{m_{\pi}^2}{m_{\omega}^2}\right)
|L_{\omega}|^2.
\end{equation}
Function $L_{\omega}$ is
$$L_{\omega}=Li_2\left(\frac{2}{1+\sqrt{\lambda_1}}\right)+Li_2\left(\frac{2}
{1-\sqrt{\lambda_1}}\right)-Li_2\left(\frac{2}{1+\sqrt{\lambda_2}}\right)
-Li_2\left(\frac{2}{1-\sqrt{\lambda_2}}\right),$$ where $\lambda_1 =
1 -4m_q^2/m_{\omega}^2, \, \lambda_2 = 1 -4m_q^2/m_{\pi}^2$ and the
decay constant $f_{\pi} = 93\,\mbox{MeV}$. In (\ref{E:5}) it was
used the Goldberger-Treiman relation $\varkappa \approx m_q/f_{\pi}$
(see, for example, \cite{24}). It defines the coupling $\varkappa$
through the constituent quark mass. Then the constituent quark mass
value, which is taken as a free parameter, can be found from the
widths fit.

The decay $\rho^0 \to \pi^0\gamma$  is described by the diagrams in
Fig.3 with the corresponding replacement of the coupling constant. The
expression for the width is
\begin{equation}\label{E:6}
\Gamma(\rho^0 \to \pi^0\gamma) =\frac{\alpha g_1^2}{3\cdot
2^7\pi^4}|c_{\phi}|^2\cdot\left(\cos\theta\cdot\frac{g_2}{g_1}\right)^2
m_q\frac{m_q^3}{m_{\rho}f_{\pi}^2}\left(1-\frac{m_{\pi}^2}{m_{\rho}^2}\right)
|L_{\rho}|^2.
\end{equation}
 Note that the relation $\Gamma(\rho^0 \to \pi^0\gamma)/\Gamma(\omega \to \pi^0\gamma) \approx 1/3^2$ follows
from the isotopic structure of $\rho qq$, $\omega qq$ and $\gamma qq$ vertices (\ref{E:12a})-(\ref{E:12b})
(here we omit small terms proportional to $\sin\theta$ and $s_{\phi}$). Both the widths are in a good agreement
with the experimental data \cite{29d} when the effective quark mass value is $m_q
= 175 \pm 5 \,\mbox{MeV}$:
\begin{align}\label{E:7}
 \Gamma^{theor}(\omega \to \pi^0\gamma)=0.74\pm
0.02 \, \mbox{MeV},\quad \quad \Gamma^{exp}(\omega
\to \pi^0\gamma)=0.76 \pm 0.02 \,\mbox{MeV};\notag \\
\Gamma^{theor}(\rho^0 \to \pi^0\gamma) = 0.081\pm 0.003 \, \mbox{MeV},
\quad \quad \Gamma^{exp}(\rho^0 \to \pi^0\gamma)=0.090 \pm 0.012 \,
\mbox{MeV}.
\end{align}
Thus, the fitting of two widths by one model parameter - mass of the constituent quark $m_q$,
gives $m_q=175 \pm 5\,\mbox{MeV}$.

This quark mass value can be understood in a connection with the schematic representation of the quark-gluon content of hadrons.

Specifically, for the nucleon and $\rho$- meson masses we can write approximately
\begin{equation}\label{E:15}
 m_N=3m_q+m_G;\,\,\,m_{\rho}=2m_q+m_G.
\end{equation}

Here we use an effective quark mass, $m_q$, and suppose the same gluon component of meson and baryon structure, $m_G$.
From this system of equations it follows $m_G\approx 435\,\mbox{MeV}$ and $m_q\approx 170\,\mbox{MeV}$. The latest value nearly coincides with
the quark mass assessment resulting from the decay channels analysis. Note that a sufficiently low  quark mass value $m_q\approx 230 \,\mbox{MeV}$
together with the $m_{\sigma}\approx 470 \, \mbox{MeV}$ are followed from the composite-meson model with the four-quark interaction ~\cite{38aa}.

So, the model discussed introduces the constituent quark which is differed from the traditional effective quark component of
a hadron with $m_q=m_N /3\approx 300\,\mbox{MeV}$ (see, however, ~\cite{38a}).

Moreover, the value of $m_G$ makes it possible to interpret the meson $\sigma_0=f_0(600)$ as
the (mostly) glueball state ~\cite{38b,38c}. It does not contradict to some known ideas on the gluon nature
of $\sigma-$ meson (see ~\cite {38} and references therein for the review of scalar meson properties).
If it is the case, the low value of the constituent quark mass can be treated as the consequence of a model separation of the quark and
gluon degrees of freedom in a hadron. Hence, the value of $m_q$, which is essentially lower than $300\,\mbox{MeV}$, provides the agreement
of the model results with the experiment for the processes under consideration.

Of course, if we explanate the $\sigma$-meson as the effective glueball state, we need in an
detailed analysis of its properties and manifestations in the model. This studying will be presented in a forthcoming paper.
Note also that the $\sigma-$ meson
was carefully analyzed, supposing the meson is a $\bar qq$ state, in ~\cite{39,40}, for instance.

In an analogy with the process $\pi\rightarrow \gamma\gamma$, the decays
under consideration can contain an anomaly contribution caused by
the $\pi qq$ vertex. This contribution is small here due to the
smallness of the ratio $m^2_{\pi}/m^2_{\rho}$, that is approximation
$m_{\pi}=0$ takes place with a good accuracy. It is known that the
same approximation is not admissible for the decay $\pi\rightarrow
\gamma\gamma$, where the anomaly contribution dominates.

An analysis of this decay in our model is coincide with the one in
~\cite{24}, for example. Namely, due to the dominant anomaly
contribution, the decay amplitude is again $$M_{\pi^0  \gamma
\gamma}=\frac{\alpha N_c}{3\pi f_{\pi}},$$ i.e. it depends on the
$f_{\pi}$ and does not contain the quark mass. The constant $f_{\pi}$ arises in the amplitude from pion-quark
coupling which is $\varkappa=m_q/f_{\pi}$, as it dictates by the
Goldberger-Treiman relation. So, our value of the quark mass
corresponds to the (lower) pion-quark coupling, $\varkappa$, providing, however,
the reasonable values for $\rho (\omega) \to \pi^0 \gamma$ widths.
 Note also that in the consideration of loop three-particle decays $\omega (\rho) \to \pi^0 \pi^+ \pi^-$
 we will have in this model a non-negligible ratio $m_{\pi}^2/ m_q^2$ as a consequence of
 this relatively low quark mass (cf. with
 ~\cite{28}). Nevertheless, an accurate calculations give for $\Gamma(\omega (\rho)\to 3\pi)$ the
 values which are in agreement with the
 data (these results will be
discussed in the forthcoming work too).

One of the most important features of the model is the implantation
of electromagnetic (EM) interaction into the gauge scheme of strong
interactions. The mixing of the gauge fields leads to the effect of
mutual renormalization of the interaction couplings. Now we discuss
the decay $\rho^0\to e^+e^-$ which is caused mainly by the EM
interaction. At the tree level this process is described by the
vertex following from (\ref{E:10}) and the mixing (\ref{E:11}).
Corresponding decay coupling is $g_{\rho ee}\approx g_1
\sin\theta=e\tan\theta$, so it gives $\Gamma^{tr}(\rho\to ee)=4.6\,
\mbox{KeV}$, while the experimental decay rate is
$\Gamma^{exp}(\rho\to ee)=7.0\,\mbox{KeV}$ \cite{29d}. Here we show
that this discrepancy can be eliminated due to renormalization of
the decay coupling $g_{\rho ee}$ by the strong interactions. In the
one-loop approximation there are two types of diagrams which give
the main contribution to the renormalization (see Fig.4). In this
figure, solid lines correspond to $\rho, \pi, q,$ and $a_0$ (or all
set of scalar particles, $s$),
 and wavy lines correspond to $\gamma$ and $\rho^0$. All extra diagrams are damped by the small mixing parameters.
\begin{figure}[h!]
\centerline{\epsfig{file=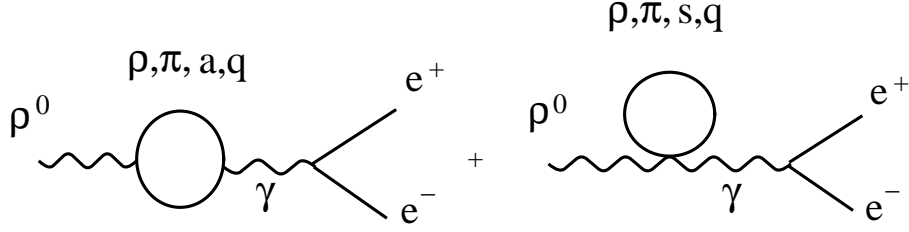,width=12cm}} \caption{Loop
diagrams for the decay $\rho^0\to e^+ e^-$.} \label{fig:Feynm3}
\end{figure}

As a result, we get full amplitude of the process:
\begin{equation}\label{E:16}
 \mathcal{M}=\mathcal{M}^{tr}\cdot [1+k\cdot\delta^{\gamma}(\mu)],
\end{equation}
where $k=(g_2^2\cos\theta /4\pi)^2$ and $\delta^{\gamma}(\mu)$ is the sum of the loop contributions.
Finally, the decay rate has the form:
\begin{equation}\label{E:17}
 \Gamma(\rho\to ee)=\kappa(\mu)\cdot\Gamma^{tr}(\rho\to ee),
\end{equation}
where $\kappa(\mu)=(1+k\cdot\delta^{\gamma}(\mu))^2$
and $\mu$ is the renormalization parameter. The function
$\delta^{\gamma}(\mu)$ in the
$\overline{MS}$ scheme is:
\begin{align}\label{E:18}
 \delta^{\gamma}(\mu)=&-\frac{1891}{108}+\frac{44}{9}k_q-\frac{7}{3}k_a
 -\frac{8}{3}k_{\pi}+\frac{k_s}{\cos^2\theta}+\frac{23}{4}\ln\mu_{\rho}
 -\frac{1}{3}(1+8k_{\pi})\ln\mu_{\pi}\notag\\&+\frac{1}{6}(6k_a-1)\ln\mu_a-
 \frac{4}{3}(1+\frac{1}{3}k_q)\ln\mu_q-\frac{k_s}{\cos^2\theta}\ln\mu_s+\frac{\sqrt{3}\cdot33}{2}\arctan\frac{1}{\sqrt{3}}\notag\\
 &-\frac{1}{3}(1-4k_a)\sqrt{\beta_a}\arctan\frac{1}{\sqrt{\beta_a}}
 -\frac{4}{3}(1+\frac{1}{2}k_q)\sqrt{\beta_q}\ln|\frac{1+\sqrt{\beta_q}}{1-\sqrt{\beta_q}}|\notag\\
 &-\frac{1}{3}(1-4k_{\pi})\sqrt{\beta_{\pi}}\ln|\frac{1+\sqrt{\beta_{\pi}}}
 {1-\sqrt{\beta_{\pi}}}|\,,
\end{align}
Here we use normalized quantities:
\begin{align}\label{E:19}
 &k_q=\frac{m^2_q}{m^2_{\rho}},\,\,\,k_a=\frac{m^2_a}{m^2_{\rho}},\,\,\,
 k_{\pi}=\frac{m^2_{\pi}}{m^2_{\rho}},\,\,\,k_s=\frac{m^2_s}{m^2_{\rho}},\notag\\
 &\mu_{\rho}=\frac{m^2_{\rho}}{\mu^2},\,\,\,\mu_{q}=\frac{m^2_{q}}{\mu^2},\,\,\,
  \mu_{a}=\frac{m^2_{a}}{\mu^2},\,\,\,\mu_{\pi}=\frac{m^2_{\pi}}{\mu^2},
  \,\,\,\mu_s=\frac{m^2_s}{\mu^2},\notag\\
 &\beta_q=1-4 k_q,\,\,\,\beta_{\pi}=1-4 k_{\pi},\,\,\,\beta_a=4 k_a-1.
\end{align}
In the MS scheme an additional term arises in (\ref{E:18}):
\begin{equation}\label{E:20}
 \Delta\delta^{\gamma}=(C+\ln\pi)(\frac{47}{12}-\frac{4}{9}k_q+k_a)\,,
\end{equation}
where $C+\ln\pi\approx 1.7219$. With the help of the expressions (\ref{E:16})-(\ref{E:20})
we have found that
$\Gamma^{th}=\Gamma^{exp}$ when the renormalization parameter
 $\mu\approx 550\,\mbox{MeV}$ in the $\overline{MS}$ scheme
 and $\mu\approx 1325\, \mbox{MeV}$ in the MS scheme.
 So, strong interactions lead to the large renormalization of the EM coupling
 and the physical result $\kappa(\mu)\approx 1.5$ is obtained within the
 framework of $\overline{MS}$ scheme when the renormalization
 scale $\mu$ is in the range of typical meson masses.
 The value of $\kappa (\mu)$ depends on the masses of quarks
 and scalar mesons very weakly. The quark mass changing
 from $175 \,\mbox{MeV}$ to $300 \,\mbox{MeV}$ leads to
 the variation of $\mu$ lower than $5 \%$. Variation of $\mu$ due to the
 increasing of scalar mass from $450 \,\mbox{MeV}$ up to $1000 \,\mbox{MeV}$ is approximately $3 \%$.
An analogous effect takes place for the decay $\omega\to e^+e^-$, but in this case we
have use free parameter ($\varepsilon$ from (\ref{E:10})).
The presence of this parameter makes it possible to get the accordance
$\Gamma^{th}(\omega\to e^+e^-)=\Gamma^{exp}(\omega\to e^+e^-)$ for the same values of $\mu$.

\section{Conclusions}

The linear sigma-model is the most popular approach in the
description of low-energy hadron interaction. We have considered
the gauge generalization of this model and include quark degrees
of freedom explicitly. In the gauge scheme the vector meson
dominance occurs in the tree-level processes. The
quark-meson sector describes some loop-level processes, when quark
structure of mesons plays a noticeable role.

The model is applied to some radiative decays of mesons which are
intensively discussed in the literature. Decays
$\rho^0\rightarrow\pi^+\pi^-\gamma$ and
$\omega\rightarrow\pi^+\pi^-\gamma$ have been considered within the
framework of the vector dominance gauge variant. The results are in
good accordance with the experimental data. The radiative decays
$\rho^0\rightarrow\pi^0\gamma$ and $\omega\rightarrow \pi^0\gamma$
take place at the loop level only, when the quark-meson interaction
is important. For these cases we get results in agreement with the
experiment too. Decay channel $\rho^0\to e^+e^-$ can be described by
the model when we take into account the electromagnetic decay
constant renormalization by strong interactions.

From the analysis we fulfilled, it follows that the radiative decay
rates of vector mesons can be calculated with a good accuracy within
the framework of the gauge model. Quantum field approach to the vector
dominance in meson interaction can be also supplemented by the
quark-meson interaction which expands the linear sigma-model.

\end{document}